\documentclass[
 reprint, 
 cha, 
 amsmath,
 amssymb,
 aip,
 groupedaddress, 
 citeautoscript, 
]{revtex4-1}

 \usepackage{dcolumn} 
 \usepackage{bm} 
 \usepackage{graphicx,color}
\usepackage{xcolor}
 
 \usepackage{hyperref}
\hypersetup{
    urlbordercolor=black, 
  colorlinks   = true, 
  urlcolor     = blue, 
  linkcolor    = blue, 
  citecolor   = blue 
}

\begin{document}

\title{Surrogate-assisted network analysis of nonlinear time series}
\author{Ingo Laut}
\email{ingo.laut@dlr.de}
\affiliation{Deutsches Zentrum f\"{u}r Luft- und Raumfahrt, Forschungsgruppe Komplexe Plasmen, 82234 We{\ss}ling, Germany}

\author{Christoph R\"{a}th}
\email{christoph.raeth@dlr.de}
\affiliation{Deutsches Zentrum f\"{u}r Luft- und Raumfahrt, Forschungsgruppe Komplexe Plasmen, 82234 We{\ss}ling, Germany}
\date{\today}

\begin{abstract}
{
The performance of recurrence networks and symbolic networks to detect weak nonlinearities in time series is compared to the nonlinear prediction error. For the synthetic data of the Lorenz system, the network measures show a comparable performance. In the case of relatively short and noisy real-world data from active galactic nuclei, the nonlinear prediction error yields more robust results than the network measures. The tests are based on surrogate data sets. The correlations in the Fourier phases of data sets from some surrogate generating algorithms are also examined. The phase correlations are shown to have an impact on the performance of the tests for nonlinearity.
}
\end{abstract}

\maketitle

\begin{quotation}
Networks generated from time series have recently attracted much attention as they provide additional information about the data under consideration. Various different ways of creating such a network from a time series have been proposed. In this paper, we use recurrence and symbolic networks to detect weak nonlinearities. The performance of these network measures is compared to the well-known nonlinear prediction error. Since often only a limited amount of data is available, we focus on relatively short time series with a few thousand time steps. We find that while all methods perform equally well for the synthetic data of the Lorenz system, the nonlinear prediction error yields the most robust results for real-world data from active galactic nuclei. The measurements are based on \emph{surrogates} which are ersatz data that contain no nonlinearities. Analyzing different surrogate generating algorithms, we find correlations in the Fourier phases of some classes of surrogates which reveal the existence of induced nonlinearities in the supposedly purely linear data sets. We show that these nonlinearities are responsible for the weak performance of the surrogates in question.
\end{quotation}

\section{Introduction}
\label{sec:introduction}
A recent milestone in the field of statistical physics has been complex network theory \cite{albert2002}. The constituents of complex systems are translated into the nodes of a network and their interactions are represented as edges. The network then contains extensive information about the system. While this procedure is straightforward for systems like social or neural networks, there is no ``natural'' way of how to create a network from a time series. 

Nonlinear time series analysis \cite{kantz2004, strogatz2006} deals with the question whether a time series has underlying chaotic dynamics. To this end, some measure of nonlinearity, most of which are derived from chaos theory, is calculated for the time series. In order to make a significant statement, the measure may also be applied to a set of so-called surrogate data sets which mimic the linear properties of the original data \cite{theiler1992}. As it is impossible to perfectly reproduce both the autocorrelation function as well as amplitude distribution, the available surrogate generation algorithms focus on different aspects \cite{theiler1992, schreiber1996, rath2012}. 

One method of creating a network from a time series is the \emph{recurrence network} \cite{xu2008, marwan2009, donner2010} motivated from recurrence quantification analysis~(RQA) \cite{eckmann1987}. In RQA, the time series is embedded in an artificial phase space. The recurrence matrix then contains the information which points are sufficiently close to each other in phase space. The structural properties of the recurrence matrix can be used to characterize different dynamic aspects of the time series \cite{eckmann1987}. By interpreting the recurrence matrix as an adjacency matrix, a network can be constructed from the time series. This approach may be used to characterize the underlying dynamical system \cite{donner2010} or detect dynamic changes by a sliding-window technique \cite{marwan2009}. 

Another approach is a network derived from an ordinal partition of the time series \cite{small2013}. In a sliding window scheme, the ordinal pattern of the windowed sequence corresponds to the one node of the network. Nodes of consecutive sequences in the time series are connected in the network in order to save the temporal information. Since the amplitude information is neglected, this approach may be combined with a transition network, where the nodes of the network are the binned amplitudes of the time series \cite{sun2014}. A node then represents a combination of amplitude binning number and ordinal pattern of the windowed sequence. 

Often it is sufficient to analyze the characteristics of the network constructed from a time series. For example, in Refs.~\onlinecite{marwan2009} and~\onlinecite{sun2014} the bifurcation diagram of a nonlinear system was analyzed. The different regimes were identified by dynamical changes of the measures as the control parameter was varied. In other cases, the network measures are not as descriptive and have to be compared to other data sets. 
In Ref.~\onlinecite{donges2011}, the network of a time series was compared to a network that was created from points which were randomly drawn from the time series. Surrogates are another means to produce ersatz data. They have the same linear properties, i.e., the same autocorrelation function, as the original data while the nonlinear properties are randomized \cite{theiler1992}. They provide a significant test for a given measure of nonlinearity by comparing the measure of the original time series to those of the surrogate data sets. 

In this paper, we use recurrence and symbolic networks to test for weak nonlinearities in time series. To this end, network measures are calculated for the time series under study and for surrogate data sets. The performance of the network tests is compared to the nonlinear prediction error. The Lorenz system is used as a source of low-dimensional chaotic time series. In order to test both the ability to detect nonlinearities and the susceptibility to erroneously do so, the time series are mixed with a linear autoregressive process. As real-world data, light curves of active galactic nuclei (AGN) are examined. We attribute the performance differences of some surrogate generation algorithms to spurious nonlinearities that are introduced during the creation of the surrogates.

The paper is organized as follows. In Sec.~\ref{sec_methods}, we present the method of testing for nonlinearities with networks. In Sec.~\ref{sec:data_sets}, we describe the data sets and the process of mixing linear and nonlinear time series. In Sec.~\ref{sec:implementation_and_results}, we compare the performance of the tests for different data sets and surrogate generating algorithms. Induced correlations in the Fourier phases of some surrogates are also examined. Finally, in Sec.~\ref{sec:discussion}, we conclude with a discussion and a summary of the main results.

\section{Methods}
\label{sec_methods}

\subsection{Networks}

\begin{figure}
\includegraphics[width=\columnwidth]{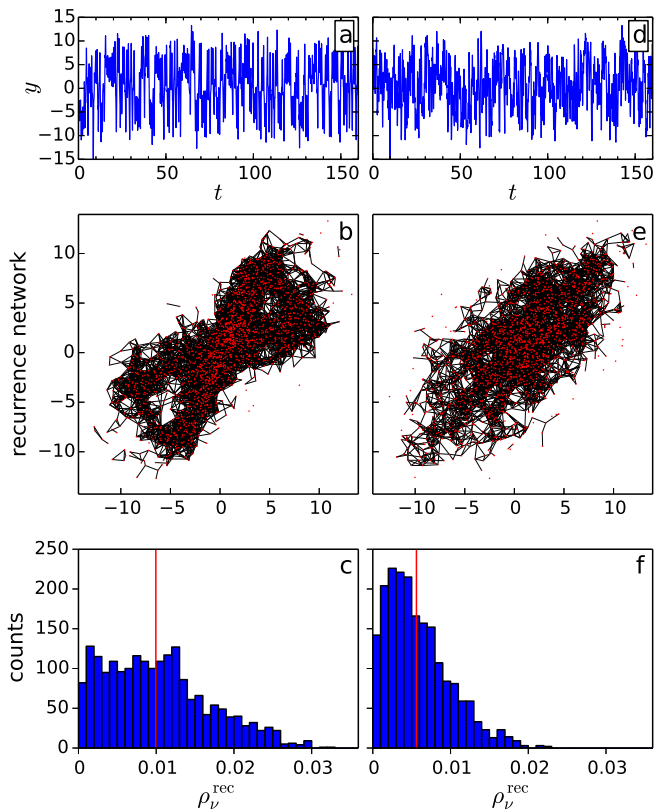} 
\caption{Comparing the recurrence networks of a nonlinear time series and its AAFT surrogate. 
(a) A nonlinear time series with mixing parameter $m=0.60$ (see Eq.(\ref{eq_mixed_lorenz})) consisting of $N=2000$ time steps with a stepsize of $\delta t = 0.08$. 
(b) The recurrence network connects nodes that are sufficiently close in the embedded phase space. The threshold $\epsilon=1.632$ was chosen such that the average connectivity obeys $\rho = 0.01$. 
The coordinates of the nodes in this representation are identical to the first two dimensions of the embedded time series into $d=3$ dimension with delay time $\tau=2 \delta t$. The butterfly-shape of the attractor of the Lorenz system is clearly visible.  
(c) The distribution of the connectivity $\rho_\nu^\text{rec}$ for the recurrence network. The average connectivity $\rho^\text{rec}$ is marked by a vertical line. 
(d)--(f) The same for an AAFT surrogate of the time series. The recurrence network in (e) is obtained for the same value of $\epsilon$ as for the original time series. 
}
\label{fig_lorenz_netw}
\end{figure}

To create a recurrence network, a time series $\{y_n\}$ of length $N$ is first embedded in an artificial phase space using the method of delay coordinates \cite{packard1980, takens1981}. For an embedding dimension~$d$ and delay time~$\tau$ the method yields the state vector $\mathbf{y}_n = (y_{n-(d-1)\tau}, y_{n-(d-2)\tau}, \dots, y_{n})$ for the time steps $n = (d-1)\tau, (d-1)\tau+1, ..., N-1$. The adjacency matrix $A$ of the recurrence network is defined as
\begin{equation}
\label{eq_adj_matrix_cyl}
A_{ij}(\epsilon) = \Theta\left(\epsilon - \left| \mathbf{y}_i - \mathbf{y}_j  \right|\right)- \delta_{ij},
\end{equation}
where $\Theta(\cdot)$ is the Heaviside function and $\epsilon$ an appropriate threshold. The adjacency matrix connects points in phase space that are sufficiently close to each other. The recurrence network has thus a total of $N_w = N - (d-1)\tau$ nodes. It is an undirected network since the adjacency matrix is symmetric. The recurrence network of the nonlinear time series shown in Fig.~\ref{fig_lorenz_netw}(a) can be seen in Fig.~\ref{fig_lorenz_netw}(b). 

The symbolic network proposed in Ref.~\onlinecite{sun2014} is a directed network. Each time step $n$ is associated with a symbol-pair containing the amplitude information $\alpha(n)$ and the ordinal pattern $\pi(n)$. The former is calculated by binning  the time series in the interval $[\min(\{y_n\}), \max(\{y_n\})]$ into $Q$ equal regions. $\alpha(n)$ is then simply the bin number of $y_n$. To compute $\pi(n)$, one considers the sequence $[y_n, y_{n+\tau}, ..., y_{n+(L-1)\tau}]$ for a given time-delay $\tau$ and window length~$L$. The ordinal pattern $\pi(n) = (\beta_1, \beta_2, ..., \beta_L)$ contains the indices sorting the sequence such that $y_{n-1+\beta_1} \leq  y_{n-1+\beta_2} \leq ... \leq y_{n-1+\beta_L}$. For example, the ordinal pattern corresponding to the sequence [1.2, 3.0, 2.0] is $\pi = (1, 3, 2)$. The symbol-pair at step $n$, ($\alpha(n)$, $\pi(n)$), is then one node of the network, and it is connected by a directed link to the symbol-pair ($\alpha(n+1)$, $\pi(n+1)$) of the successive time step. In this representation, different time steps may correspond to the same symbol-pairs. Only nodes whose symbol-pairs are present in the time series are part of the network, the actual number of nodes may thus be smaller than the maximal possible value $N_w^\text{max}=Q\cdot L!$.

It was argued that the symbolic network provides a simple and robust scheme for phase space repartition where the window length $L$ plays the role of the dimension $d$ for time delay embedding \cite{small2013, sun2014}.  In Fig.~\ref{fig_lorenz_sunnet}(a), the symbolic network constructed from the nonlinear time series of Fig.~\ref{fig_lorenz_netw}(a) is shown.

\subsection{Measures}


For the recurrence network, which is an undirected network, the local connectivity $\rho^\text{rec}_\nu$ is calculated by normalizing the number of nodes that are connected to node $\nu$ by the maximal value \cite{donner2010}, 
\begin{equation}
\label{eq_local_connectivity}
\rho_\nu^\text{rec} =  \frac{1}{N_w-1} \sum_{i} A_{\nu, i}, ~~ \rho^\text{rec} = \frac{1}{N_w} \sum_\nu \rho_\nu^\text{rec}.
\end{equation}
Here, $N_w$ is the number of nodes of the network and $A_{\nu, i}$ is the adjacency matrix of a recurrence network. The average connectivity $\rho^\text{rec}$ is calculated by averaging over all nodes of the network. If the attractor of the nonlinear time series is successfully reconstructed by the embedding, the recurrent trajectories will lead to a larger value of $\rho^\text{rec}$ as compared to linear data sets. This can be seen in Fig.~\ref{fig_lorenz_netw}, where the average connectivity $\rho^\text{rec}$ of a recurrence network derived from a nonlinear time series is compared to $\rho^\text{rec}$ from surrogate data where the nonlinearities have been removed.

\begin{figure}
\includegraphics[width=\columnwidth]{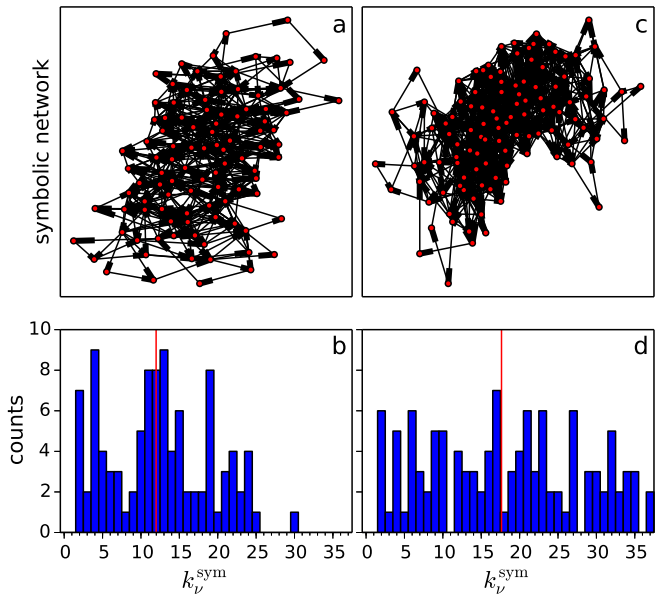} 
\caption{Comparing the symbolic networks of the nonlinear time series of Fig.~\ref{fig_lorenz_netw} and its AAFT surrogate. 
(a) The nodes (circles) of the network consist of symbols which are associated to each step of the time series, containing the ordinal pattern and amplitude information with $L=3$, $\tau=2 \delta t$ and $Q=25$. The symbols of two successive time steps are connected by directed, unweighted links whose end points are indicated by a thicker end.
(b) The corresponding degree distribution. The average degree $k^\text{sym}$ is marked by a vertical line. 
(c)--(d) The same for the AAFT surrogate of the time series. }
\label{fig_lorenz_sunnet}
\end{figure}

For the directed symbolic network described above, the degree $k^\text{sym}_\nu$ counts the number of links ending and starting at node $\nu$ separately,
\begin{equation}
\label{eq_degree_directed}
k^\text{sym}_\nu = \sum_i A_{\nu, i} + \sum_i A_{i, \nu}, ~~ k^\text{sym} = \frac{1}{N_w} \sum_{\nu} k^\text{sym}_\nu.
\end{equation}
As can be seen in Fig.~\ref{fig_lorenz_sunnet}, the nonlinear time series has a smaller value of the average degree $k^\text{sym}$ than the linear surrogate. This can be understood by noting that the quasi-periodic orbits (QPOs) of the attractor lead to identical links between the symbols of the network, and thus to a smaller average degree.

The nonlinear prediction error (NLPE) is a commonly used test for nonlinearity \cite{sugihara1990, kantz2004}. In Ref.~\onlinecite{schreiber1997} it was found to be the one with the best overall performance for a broad range of applications. The NLPE of an embedded time series \{${\mathbf{y}_n}$\} is defined as \cite{sugihara1990}
\begin{equation}
\label{eq_nlpe}
\mathcal{E}= \frac{1}{\sqrt{N-T}} \sqrt{ \sum_{i=0}^{N - T -1} \left( \mathbf{y}_{i+T} - \mathbf{F}[\mathbf{y}_i, g] \right)^2 },
\end{equation}
where $T$ is the lead time and $\mathbf{F}$ is a predictor. The predictor $\mathbf{F}$ is calculated by averaging over the future values of the $g$ nearest neighbors of point $\mathbf{y}_i$ a lead time $T$ ahead. As the NLPE is calculated in an artificial embedding space, it also implicitly depends on the embedding dimension $d$ and the delay time $\tau$. 

In order to perform a statistical test, the measures are compared to surrogate data described below. The size $\alpha$ of a test is the probability that the null hypothesis is rejected, although it is in fact true \cite{schreiber1997}. A measure $M$ is compared to the measures $\{M_\text{surro}\}$ of $B$ realizations of the surrogate data. As one expects a larger value of the average connectivity for recurrence networks in the presence of nonlinearity, the null hypothesis is rejected if $\rho^\text{rec}$ is larger than all the $\{ \rho^\text{rec}_\text{surro} \}$. The number of surrogates needed to achieve a given size of this \emph{one-sided} test is $B=1/\alpha - 1$. In the same manner, as one expects \emph{smaller} values of the average degree for symbolic networks in the presence of nonlinearity, the null hypothesis is rejected if $k^\text{sym}$ is smaller than the $\{ k^\text{sym}_\text{surro} \}$. In the case of the NLPE, the null hypothesis of linearity is rejected if $\mathcal{E}$ is smaller than all the $\{ \mathcal{E}_\text{surro} \}$. By repeating the test, one can calculate the discrimination power $D(M)$ as the ratio of the number of rejections of the null hypothesis to the total number of tests.

For real-world data there are often not enough time series available to calculate the discrimination power. In this case, the significance of the deviation of the time series from a set of surrogates is calculated as $S(M) = \left|M - \langle \{M_\text{surro}\} \rangle \right| / \sigma(\{M_\text{surro}\})$.

\subsection{Surrogate algorithms}

Surrogates are an important tool for the detection of nonlinearities in time series \cite{theiler1992, schreiber1996, ivanov2004, blinowska2004, bassett2013}. They are data sets which mimic the linear properties, i.e., the autocorrelation function, of the original data while possible higher order correlations are randomized. The most commonly used methods for generating surrogates are Fourier transformed (FT) surrogates and their  amplitude-adjusted (AAFT) and iterative amplitude-adjusted (IAAFT) generalizations.

FT surrogates are compatible with the null hypothesis of a linear Gaussian process \cite{theiler1992}. They are generated by randomizing the phases of the discrete Fourier transform of the original time series and subsequently performing the inverse transform. The Wiener-Khinchin theorem guarantees the surrogates to have the same autocorrelation function as the original time series. Being truly linear, the surrogates can unveil higher order correlations, however, this test is limited to time series which themselves obey a Gaussian distribution. The original time series therefore has to be rank-ordered-remapped to a Gaussian distribution prior to the analysis \footnote{By rank-ordered remapping a data set $\protect
  \{x_n$\protect \} to $\protect \{y_n\protect \}$, we understand a reordering
  of $\protect \{y_n\protect \}$ such that if $x_i$ is the $i$th smallest of
  the $\protect \{x_n\protect \}$, then $y_i$ is the $i$th smallest of the
  $\protect \{y_n\protect \}$. The reordered $\protect \{y_n\protect \}$ thus
  ``follow'' the original set $\protect \{x_n\protect \}$ while having an
  amplitude distribution identical to the one of $\protect \{y_n\protect
  \}$.}.

AAFT surrogates extend the null hypothesis to a Gaussian process which was distorted by an instantaneous, time-independent measurement function \cite{theiler1992}. Here, a copy of the original time series is first rank-ordered remapped  to a set of Gaussian random numbers. Then, FT surrogates of this remapped time series are created. Finally, the surrogates are rank-ordered remapped to the original time series. The surrogate now mimics both the autocorrelation function and the amplitude distribution. The final step, however, leads to a whitening of the power spectrum as compared to the original time series. 
In Ref.~\onlinecite{schreiber1996} it was shown that this may lead to an erroneous detection of nonlinearity in purely linear time series. 

IAAFT surrogates were designed to overcome this shortcoming \cite{schreiber1996}. The method for generating IAAFT surrogates starts with a random shuffle $\{s_n\}$ of the original time series $\{y_n\}$. The Fourier amplitudes $\{\Phi_n\}$ of the original time series are saved. Now, the following two steps are repeated iteratively. (1) Take the Fourier transform of $\{s_n\}$, replace the corresponding Fourier amplitudes by $\{\Phi_n\}$, and transform back. $\{s_n\}$ has now exactly the same autocorrelation function as $\{y_n\}$, but not the same amplitude distribution. (2) Rank-ordered remap the resulting $\{s_n\}$ to $\{y_n\}$. As step (2) changes the power spectrum of the surrogate $\{s_n\}$ the two steps are repeated until the rank-ordered remapping no longer leads to a change in the surrogate.

\begin{figure*}
\centering
\includegraphics[width=\linewidth]{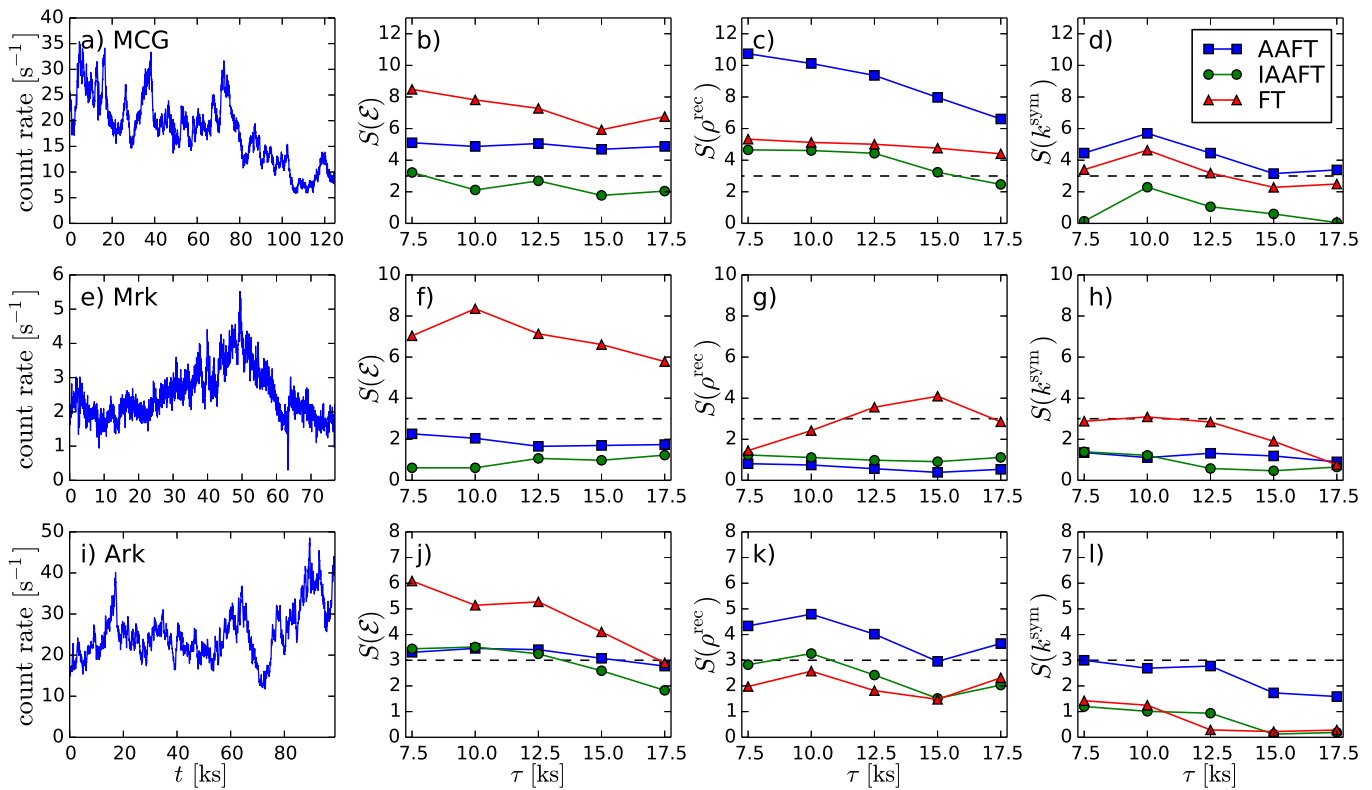}
\caption{Detection of nonlinearities in AGN time series. 
(a) The MCG time series, consisting of $N = 2497$ steps. 
(b) The significances $S(\mathcal{E})$ of the nonlinear prediction error $\mathcal{E}$ as a function of delay time $\tau$. 
(c) The significances $S(\rho^\mathrm{rec})$ of the average connectivity $\rho^\text{rec}$ for recurrence networks as a function of $\tau$. 
(d) $S(k^\mathrm{sym})$ for symbolic networks as a function of $\tau$. 
(e)--(h) The same for the Mrk time series consisting of $N = 1540$ steps.
(i)--(l) The same for the Ark time series of $N = 1978$ steps.
The step length of all time series is $\delta t = 50~\mathrm{s}$. To calculate the significance, 400 surrogates were generated for each of the surrogate generating algorithms FT, AAFT and IAAFT. For each measure, the $3\sigma$ detection limit is shown as a dashed line. }
\label{fig_agn_sign}
\end{figure*}

\section{data sets}
\label{sec:data_sets}
The tests for nonlinearity are applied to the well known Lorenz equations \cite{lorenz1963}. Time series containing $N = 2000$ data points with a stepsize of $\delta t = 0.08$ time units are considered (similar to Ref.~\onlinecite{schreiber1997}). The $x$ coordinates $\{x_n\}$ of the system are mixed with a (linear) autoregressive (AR) process $\{a_n\}$ in order to generate a new series $\{y_n\}$ that contains a fraction $m$ of the nonlinear series. The same distorted AR process as in Ref.~\onlinecite{schreiber1996} is used, it reads $a_n = s_n \sqrt{|s_n|}, ~s_n = cs_{n-1} + \eta_n$ with $c=0.9$ and noise $\eta_n$ drawn from a Gaussian distribution.

The mixed time series,
\begin{equation}
\label{eq_mixed_lorenz}
y_n = m \cdot x_n + (1-m) \cdot a_n,
\end{equation}
is analyzed in order to test both the ability to detect weak nonlinearities and the susceptibility to erroneously reject the null hypothesis for linear time series. A realization of a mixed time series for $m=0.6$ and its AAFT surrogate can be seen in Figs.~\ref{fig_lorenz_netw}(a) and \ref{fig_lorenz_netw}(c). 

In astrophysics, the analysis of light curves from AGN is important as the detection of nonlinearities can be used to test different theoretical models for the energy production at the center of the host galaxy \cite{peterson1997, ulrich1997}. An AGN is a luminous region at the very center of a galaxy that is powered by accretion onto a supermassive black hole. Seyfert galaxies discussed here are a subclass of radio-quiet AGN which are again subdivided in different types ranging from 1 to~2, depending on the observed line widths. It was argued that the variety of AGN is partly the result of different aspect angles \cite{antonucci1993}. 

The contributions of a test for nonlinearity are twofold. First, linear models like global disk oscillation models \cite{titarchuk2000} can be rejected with the detection of nonlinearities. Second, the analysis of the (nonlinear) dynamics can put to test different nonlinear models. For example, it was argued that AGN are galactic black hole binary systems with their masses scaled up \cite{mchardy2006}. Black hole binaries can be in the state of quasi periodic oscillation~(QPO) \cite{remillard2006}. One evidence of the relation between galactic black hole systems and AGN would be the detection and characterization of QPOs in AGN data \cite{antonucci1993}. A unified model for these objects of very different masses and time scales would contribute greatly to the understanding of the physical processes close to a black hole. 

The AGN data were taken from the public archive of the XMM-Newton satellite. The measured count rate of the pn-CCD camera was background subtracted and binned to a stepsize of $\delta t = 50~\mathrm{s}$. The first light curve considered here is a measurement of the Seyfert galaxy {MCG-6-30-15} (MCG) taken during the 303th revolution of the satellite around the earth \cite{vaughan2003}. The time series is shown in Fig.~\ref{fig_agn_sign}, top row. With a duration of more than $120$ kiloseconds (ks), or $N = 2497$ time steps, it is the longest time series considered here.
The bright narrow-line Seyfert~1 galaxy Mrk~766 (Mrk) has been observed by all main X-ray observatories. Here, a measurement taken by the XMM-Newton satellite, revolution 999, is used \cite{markowitz2007}. The light curve is shown in Fig.~\ref{fig_agn_sign}, middle row. Power density spectra have been analyzed by Markowitz \emph{et al.}\cite{markowitz2007}, and a test for nonlinearity has previously shown a very significant outcome \cite{rath2012}. The significance of the detection of a QPO in another measurement on Mrk \cite{boller2001} was questioned by Benlloch \emph{et al}\cite{benlloch2001}. 
The XMM-Newton observation of Ark~564 (Ark) during revolution 930 is also examined \cite{arevalo2006} (see Fig.~\ref{fig_agn_sign}, bottom row).

\begin{figure}
\includegraphics[width=\columnwidth]{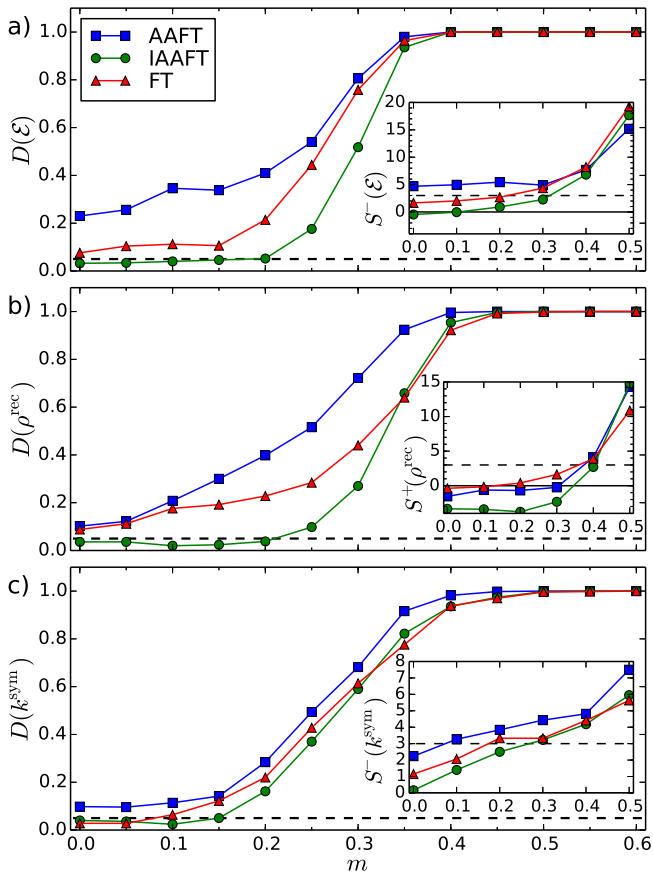}
\caption{Detection of nonlinearities in the mixed Lorenz system (see Eq.~(\ref{eq_mixed_lorenz})). 
(a) Discrimination power $D(\mathcal{E})$ of the nonlinear prediction error as a function of mixing parameter $m$. Tests were carried out for 500 realizations of the mixed Lorenz system and three surrogate generating algorithms AAFT, IAAFT and FT. The size $\alpha=0.05$ of the test is shown as a dashed line.
(b) The same for $D(\rho^\mathrm{rec})$ for the average connectivity of recurrence networks. 
(c) The same for $D(k^\mathrm{sym})$ for the average degree of symbolic networks.  For comparison, the signed significance of a single realization of the time series is shown as an inset for each measure $M$ as $S^{\pm}(M) = \pm ( M - \langle\{M_\text{surro}\}\rangle ) / \sigma(\{M_\text{surro}\}$. In the insets, the dashed line depicts the $3\sigma$ detection limit and the solid line $S^\pm = 0$. }
\label{fig_lorenz_pwr}
\end{figure}

\section{results}
\label{sec:implementation_and_results}

\subsection{Lorenz data}
The delay time for the embedding of the mixed Lorenz data is set to $\tau = 2 \delta t = 0.16$ such that the attractor is clearly reconstructed for mixing parameter $m=1$ (the ``pure'' Lorenz data). The appropriate embedding dimension $d$ is determined with the false nearest neighbor statistics \cite{kennel1992}. This method confirms the absence of false neighbors in the case of mixing parameter $m=1$ for $d=3$. To calculate the discrimination power $D(M)$ for the three measures $M$ described in Sec.~\ref{sec_methods}, the tests of size $\alpha = 0.05$ are repeated 500 times. For $m=0$ one expects $D(M) = \alpha$ as there are no nonlinearities present in the data and the null hypothesis is only rejected \emph{by chance} when comparing the linear time series to the $B = 1/\alpha - 1$ surrogates. Larger values of $D(M)$ suggest then that the measure $M$ is susceptible to erroneously detect nonlinearity in linear data. By increasing $m$, it can be studied how well weak nonlinearities are detected by the measure. Having checked that all tests perfectly detect the nonlinearities for $m > 0.6$ we show results for the values of $m$ in the range $m =0, 0.05, ..., 0.6$. 

Varying the lead time $T$, the best results for the NLPE were found for $T = 2 \delta t = 0.16$. The discrimination power $D(\mathcal{E})$ as a function of $m$ can be seen in Fig.~\ref{fig_lorenz_pwr}(a). The AAFT surrogates erroneously reject the null hypothesis at $m=0$, where the series is a pure AR process, with a probability of 23~\%. This observation was used in Ref.~\onlinecite{schreiber1996} to argue the advantages of the IAAFT algorithm which indeed shows a lower percentage of false rejections. There, however, the performance of the algorithms was not compared for time series with spurious nonlinearities. In Fig.~\ref{fig_lorenz_pwr}(a) it can be seen that at $m=0.2$, the rejection probability of the IAAFT algorithm is still 5.2~\% which is very close to the size of the test. Before using the FT algorithm, the data sets are rank-order-remapped to a Gaussian distribution. The FT algorithm shows a good performance, starting at about 7.6~\% at $m=0$ and showing a rejection probability of 21.4~\% at $m=0.2$. 

The same embedding parameters $d=3$ and $\tau = 2 \delta t$ are used for the recurrence network. The threshold~$\epsilon$ is chosen such that the average connectivity of the time series is $\rho = 0.01$. The same threshold is then used for the surrogate data. Figure \ref{fig_lorenz_pwr}(b) shows the discrimination power $D(\rho^\text{rec})$ of the test. The overall performance is similar to the NLPE, but slightly weaker. For $m=0$, the FT and the AAFT surrogates have a slightly increased rejection probability of 8.8~\% and 10.2~\%, respectively. For intermediate values of the mixing parameter, $0.1<m<0.4$, the AAFT surrogates again show a high rejection probability, while the one for the IAAFT surrogates is very low.

Since the role of the window length $L$ of the ordinal pattern is comparable to the embedding dimension $d$ of time-delay embedding, $L=3$ is used for the generation of symbolic networks. The same delay time $\tau = 2 \delta t$ is used as for the embedding with delay coordinates. The amplitudes are binned to $Q=25$ bins. The performance is again comparable to the NLPE (see Fig.~\ref{fig_lorenz_pwr}(c)). AAFT surrogates have a slightly increased rejection probability $D(k^\text{sym}) \simeq 9.8~\%$ at $m=0$, while it is again smaller for the IAAFT surrogates at intermediates values of $m$. Here, however, the difference in performance of the three surrogate generating algorithms is very small. 

Varying the embedding dimension $d$ to larger values has only a weak influence on the results of the NLPE, while the discrimination power of the recurrence networks is already noticeably reduced for $d=4$. Also the discrimination power of the symbolic networks is substantially reduced for larger $L$. Increasing the lead time $T$ of the NLPE has almost no influence on the results.

The significance of tests based on a single realization of the time series is shown in the insets of Fig.~\ref{fig_lorenz_pwr}. To facilitate the interpretation of the results for small $m$, the \emph{signed} significance is calculated as $S^{\pm}(M) = \pm ( M - \langle\{M_\text{surro}\}\rangle ) / \sigma(\{M_\text{surro}\}$ which only yields large (positive) values if the measure $M$ is larger ($S^+$) or smaller ($S^-$) than the average value of the surrogates, in conformity with the one-sided test of $D(M)$. For each test 200 surrogates were used. The AAFT surrogates yield values above or near the $3\sigma$ detection limit for the NLPE and for the symbolic networks already at $m=0$. A peculiarity is the negative value of $S^+(\rho^\text{rec}) \simeq -4$ at $m=0$ for the IAAFT surrogate which indicates that this measure detects \emph{more nonlinearity} in the surrogates than in the (linear) time series. For larger $m$, the transition to significant values is comparable to the one of the discrimination power.

\subsection{AGN light curves}
One criterion of choosing the delay time $\tau$ for the AGN data is the first zero crossing of the autocorrelation function which varies from about $150 \delta t = 7.5~\mathrm{ks}$ (Ark) to $350 \delta t = 17.5~\mathrm{ks}$ (MCG). A range of $\tau = 7.5, 10, 12.5, 15, 17.5\,$\,ks is chosen. Determining the embedding dimension $d$, one has to find a compromise between avoiding false neighbors and maintaining enough data points due to the large delay times $\tau$ necessary for the data sets. $d=3$ is a reasonable choice, as the number of false neighbors is already well below 10~\% for the AGN light curves considered here. The lead time for the NLPE is set to $T=250$\,s, or 5 time steps. The significances $S(\mathcal{E})$ of the NLPE for the three AGN curves depend strongly on the choice of the surrogate generation algorithm (see Fig.~\ref{fig_agn_sign}, second column). The FT surrogates show high significances for all curves. The AAFT surrogates show significant values for the MCG and slightly significant values $S(\mathcal{E}) \simeq 3.3$ for the Ark, while the IAAFT surrogates stay near or below the $3\sigma$ detection limit for all curves. 

The significances of the tests from the recurrence networks with $d=3$ are shown in Fig.~\ref{fig_agn_sign}, third column, as a function of delay time $\tau$. The AAFT algorithm yields high significances for the MCG ($S(\rho^\text{rec}) \simeq 10$) and Ark ($S(\rho^\text{rec}) \simeq 5$) time series, while the other two algorithms show smaller significances. For the Mrk time series, the FT surrogates yield the highest significance $S(\rho^\text{rec}) \simeq 4$, the AAFT and IAAFT surrogates yield very low significances as was already the case for the NLPE. 

To achieve significant results for the symbolic network, the sliding-window length is reduced to $L=2$. The number of bins $Q=25$ is the same as for the Lorenz data. The obtained significances $S(k^\text{sym})$ are similar to those of the recurrence networks, but smaller. The AAFT surrogates yield significant values $S(k^\text{sym}) \simeq 6$ for the MCG data, and values very close to the detection limit for the Ark data. As for the previous tests, no significant values are obtained from AAFT or IAAFT surrogates for the Mrk data. 

The results depend rather weakly on the embedding dimension~$d$. While the significances of the NLPE decrease with increasing $d$, $S(\rho^\text{rec})$ tends to slightly increase for the FT surrogates. No significant results for $S(k^\text{sym})$ were obtained for larger values of $L$. For $d$ or $L$ larger than 4, the number of data points becomes too small for a reasonable analysis. Using the \emph{signed} significance defined in the caption of Fig.~\ref{fig_lorenz_pwr} only changes the sign of some measures with insignificant values $S<1$.

In Ref.~\onlinecite{rath2012} it was shown that the rank-ordered-remapping steps of the AAFT and IAAFT algorithms can introduce correlations in the Fourier phases of the surrogates. This may lead to a nondetection of weak nonlinearities in the time series. The FT surrogates, which are not rank-ordered-remapped, are not prone to phase correlations by construction. In the following, we will examine nonlinearities in the form of phase correlations in these surrogates for the Mrk data, where the difference between the FT surrogates and the (I)AAFT surrogates was most pronounced.

\subsection{Phase correlations}
\label{sec:phase_correlations}
In order to shed more light on the differences of the surrogate generation algorithms, the Fourier phases are analyzed using phase maps \cite{chiang2002}. The phases of the Fourier modes~$\{\phi_i\}$ are plotted versus the phases $\{\phi_{i+\Delta}\}$ of the modes that where shifted by a phase shift $\Delta$. For a linear time series, the~$\{\phi_n\}$ are independent, and thus scattered uniformly in the square bounded by~$\pm \pi$. Structure in a phase map shows that the~$\{\phi_n\}$ are not independent, which means that the phases contain information about the time series. One way to quantify the correlation in the phase maps is to calculate the cross correlation between $\{\phi_i\}$ and~$\{\phi_{i+\Delta}\}$ \cite{rath2012},
\begin{equation} 
 \label{eq_crosscorrelation}
 c(\Delta) = \frac{\langle \phi_i \phi_{i+\Delta} \rangle}{\sigma(\{\phi_i\}) \sigma(\{\phi_{i+\Delta}\})} \,.
\end{equation}
The average is performed over all possible pairs of shifted phases. $c$ is normalized by the standard deviation of the phases, $\sigma(\{\phi_i\})$. 

The FT scheme, which stops after randomizing the phases of the Fourier modes, is guaranteed to contain no phase information. The rank-ordering (AAFT) and the iteration scheme (IAAFT) can reintroduce phase correlations to the surrogates that only at the beginning of the algorithms were truly linear. As was shown for AGN and financial market data in Ref.~\onlinecite{rath2012}, these nonlinearities lead to a non-detection of nonlinearities in the time series.

\begin{figure}
\includegraphics[width=\columnwidth]{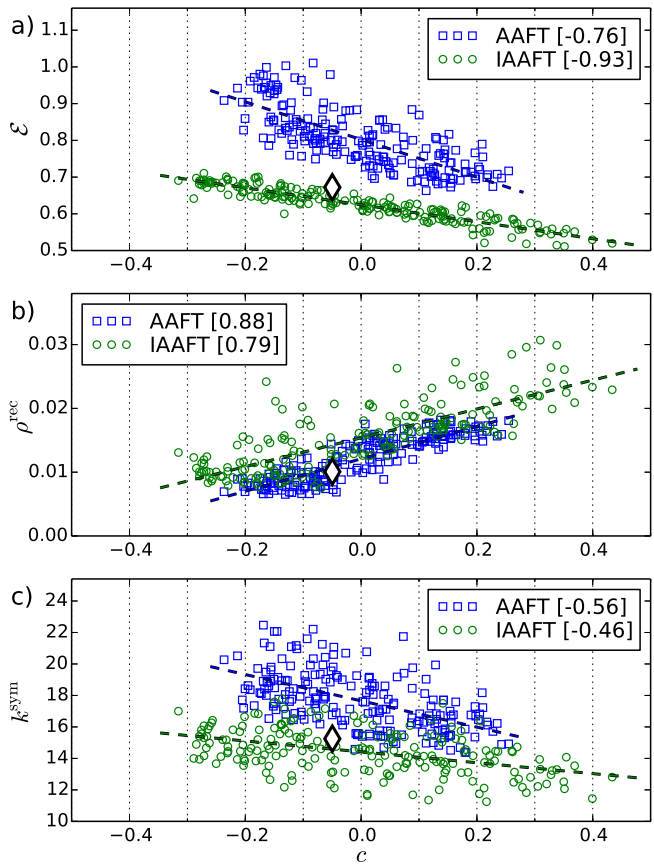} 
\caption{Phase correlations in the surrogate data of the Mrk time series.
(a) The Nonlinear prediction error $\mathcal{E}$ as a function of the cross correlation in the phase maps $c$ for $\Delta=1$ (see Eq.~(\ref{eq_crosscorrelation})). A strong anticorrelation between the measures is clearly visible, the value of the cross correlation between $\mathcal{E}$ and $c$ is given in square brackets in the legend for each surrogate generating algorithm. 
(b) The same for $\rho^\text{rec}$ calculated for recurrence networks.
(c) The same for $k^\text{sym}$ calculated for symbolic networks.
The values of the original time series are indicated by a white diamond. The dashed lines are shown to guide the eye.}
\label{fig_agn_nlpe_vs_corr}
\end{figure}

In order to examine the impact of the phase correlations on the measures for nonlinearity, the NLPE of 200 surrogates of the Mkn data is plotted in Fig.~\ref{fig_agn_nlpe_vs_corr}(a) versus the cross correlation $c(\Delta=1)$ of the phase maps. The delay time of the embedding is $\tau = 12.5$~ks. An anticorrelation for AAFT and IAAFT surrogates is clearly visible. This anticorrelation is quantified by calculating the cross correlation between $\mathcal{E}$ and $c$, the values are shown in the legend of Fig.~\ref{fig_agn_nlpe_vs_corr}. For the FT surrogates there are no significant correlations. 

Comparable results can be found for the recurrence network. In Fig.~\ref{fig_agn_nlpe_vs_corr}(b), the average connectivity $\rho^\text{rec}$ is plotted versus the cross correlation $c(\Delta=1)$. The values are now correlated instead of anticorrelated. This can be understood by noting that nonlinearities in the surrogates, quantified by a large value of $c$, lead to larger values of the average connectivity $\rho^\text{rec}$. In the case of the symbolic networks, where the difference between FT and (I)AAFT surrogates was less pronounced, $k^\text{sym}$ and $c$ are again anticorrelated, but the magnitude of the correlation coefficient is not as large as for the other two measures (see Fig.~\ref{fig_agn_nlpe_vs_corr}(c)). 

\section{discussion}
\label{sec:discussion}
The measures for nonlinearity derived from networks show a performance comparable to the NLPE when analyzing synthetic data of the mixed Lorenz equations. The different surrogate generating algorithms show the same tendencies for all measures: The AAFT surrogates tend to have a larger rejection probability, which may lead to a false rejection of the null hypothesis, i.e., a detection of nonlinearity in a linear data set. The IAAFT surrogates that were designed to overcome this problem show a relatively small rejection probability even when there is already a significant nonlinearity present in the data. Such a nondetection of nonlinearity may lead to a wrong (linear) modeling of the (nonlinear) system. The FT surrogates show the expected increase of the rejection probability when the nonlinearity is increased.

In the case of AGN data, the differences are more severe. Significant values for all three light curves considered here are only obtained for the nonlinear prediction error in combination with FT surrogates. The analysis confirms that this measure is among the most robust measures for nonlinearity \cite{schreiber1997} and also that FT surrogates are not prone to induced nonlinearities during their generation \cite{rath2012}. The results suggest that the recurrence network is better suited to reconstruct the underlying phase space than the symbolic network which in all cases yields smaller significances. The tests based on networks yield significant values with AAFT surrogates only for two of the three light curves considered here. 

In the special case of the Mrk light curve, an analysis of the phase relations of the AAFT and IAAFT surrogates shows that these algorithms induce phase correlations in the surrogate data which are correlated with the outcome of the test for nonlinearity \cite{rath2012}. These nonlinearities can explain the great differences between the significances of the FT surrogates on the one hand, and the AAFT and IAAFT surrogates on the other hand. 

Analyzing the information encoded in the Fourier phases provides additional insight in the nonlinearities of a time series. In Ref.~\onlinecite{rath2015}, the inverse approach was performed by analyzing the impact of manually created phase correlations on the outcome of nonlinearity measures. The correlations in the Fourier phases may thus not only help for checking surrogates for their linearity, but also provide a powerful tool to discriminate linear from nonlinear time series \cite{schreiber2016}. 

The analysis showed that surrogates provide a model-independent statistical test for measuring nonlinearities also in the rapidly evolving field of network analysis. Using FT surrogates of data that was rank-order-remapped to a Gaussian profile (or already obey a Gaussian distribution) seems to be the safest choice. In Ref.~\onlinecite{charakopoulos2014}, network measures derived from time series were compared to those of random networks \cite{erdos1960}. It is probable that the outcome of the test depends on the type of random network used. ``Surrogate networks'' have been also been proposed \cite{ansmann2011, mastrandrea2014}.

To conclude, we have compared tests for nonlinearity derived from networks to the nonlinear prediction error. While the performances are similar for the mixed Lorenz data, the nonlinear prediction error in combination with FT surrogates yields the most robust results for the real-world AGN data. Recurrence networks in combination with AAFT surrogates yield significant values only in two of the three AGN curves considered here. In the analysis of the third curve, we found phase correlations in the AAFT and IAAFT surrogates, which lead to a nondetection of the nonlinearities. Similar but less significant results are found for symbolic networks. 

\begin{acknowledgments}
The authors would like to thank the anonymous reviewers for their valuable comments that improved the quality of the paper.
This work has made use of observations obtained with XMM-Newton, an ESA science mission with instruments and contributions directly funded by ESA member states and the U.S. (NASA).
\end{acknowledgments}

\bibliography{./../literature}

\end{document}